\documentclass[preprint,12pt]{aastex}

\def\lsim {$\rlap{\raise.4ex\hbox{$<$}}\lower.55ex\hbox{$\sim$}\,$}

\def\c17o{$\rm C^{17}O$}
\def\dc18o{$\rm C^{18}O$}

\begin{document}

\title {\bf The Solar Nebula on Fire: A Solution to the Carbon Deficit in the Inner Solar System }
\author {Jeong-Eun Lee\altaffilmark{1},
Edwin A. Bergin\altaffilmark{2},
Hideko Nomura\altaffilmark{3}}
\altaffiltext{1}{Department of Astronomy and Space Science, Astrophysical
Research Center for the Structure and Evolution of the Cosmos,
Sejong University, Seoul 143-747, Korea; jelee@sejong.ac.kr}
\altaffiltext{2}{Department of Astronomy, The University of Michigan, 500 Church Street, Ann Arbor, Michigan 48109-1042; ebergin@umich.edu}
\altaffiltext{3}{3Department of Astronomy, Graduate School of Science, Kyoto University, Kyoto 606-8502, Japan; nomura@kusastro.kyoto-u.ac.jp}

\begin{abstract}
Despite a surface dominated by carbon-based life, the bulk composition of the Earth is dramatically carbon poor when compared to the material available at formation.  Bulk carbon deficiency extends into the asteroid belt representing a fossil record of the conditions under which planets are born.   The initial steps of planet formation involve the growth of primitive sub-micron silicate and carbon grains in the Solar Nebula.   We present a solution wherein primordial carbon grains are preferentially destroyed by oxygen atoms ignited by heating due to stellar accretion at radii $< 5$ AU.   This solution can account for the bulk carbon deficiency in the Earth and meteorites, the compositional gradient within the asteroid belt, and for growing evidence for similar carbon deficiency in rocks surrounding other stars.
\end{abstract}


\section{Introduction}

The early stages of planet formation rely on the growth of small grains into larger bodies.    Locally, the relative chemical composition of planets, comets and asteroids within the solar system provides indirect clues to the environment and processes that existed and operated billions of years ago.
In this regard, the bulk elemental composition represents a key fossil record, as it can be compared to the distribution of elements available at birth as represented by the Sun, interstellar clouds, or extra-solar protoplanetary disks.   

 Figure 1 illustrates the dramatic gradient in the amount of carbon relative to silicon incorporated into planets, comets, and asteroids, represented by meteorites.  In this figure the H/Si ratio serves to separate the various bodies with the proviso that the Sun/ISM represents the total amount of material available at the start.  Material that formed far from the Sun (comets) incorporate water and other hydrogen rich ices, but do not include the H$_2$ gas that is the major component of the Sun and ISM.  In contrast rocks within the terrestrial planet region (r $< 5$ AU) incorporate successively less H-rich ices, a well known result due to the snow-line lying inside the asteroid belt.  Icy bodies have an abundance of carbon in the form of ices and amorphous carbon grains.  In contrast the Earth has three orders of magnitude less carbon (relative to Si) than was available during formation as represented by the Sun/ISM.  While the Earth measurement does have uncertainty, as we do not know the amount of carbon sequestered in the core, estimates of the amount of hidden carbon still lead to significant carbon deficit (All\'egre et al. 2001).  Moreover, primitive carbonaceous meteorites, as tracers of the starting materials for the Earth, also exhibit carbon deficiency.   
 
In ISM $\sim 60$\% of the cosmic carbon is locked in some form of graphite or amorphous carbon grains (Savage \& Sembach 1996). These primordial grains need to be destroyed prior to planet formation to account for the composition of the Earth and carbonaceous chondrites.  That carbon grains have been detected in comets with near-solar abundances (Fig. 1) suggests that any destruction mechanism was inactive in the cometary formation zone near the giant planets.   Another hint lies in the variation of spectral classes within the asteroid belt, which may be indicative of compositional gradients in the relative amounts of carbonaceous material  (Gradie \& Tedesco 1982).  Thus there are hints in the solar system of processes that were active in the hot inner nebula that kept primordial carbon grains from being incorporated into rocks.  However, the same process was inactive in the colder outer nebula.  

In addition, the detection of carbon and iron in the atmosphere of some white dwarfs (Jura 2006) is an unexpected result as the strong gravitational field of a white dwarf is anticipated to lead to settling of heavy elements in a photosphere dominated by either hydrogen or helium (Paquette et al. 1986). The measured C/Fe abundance ratio in the atmosphere of these white dwarfs is well below solar and consistent with external supply by rocky bodies that are deficient in carbon (Jura 2008).  In sum, the solution to the carbon deficit in the inner solar system is a natural outcome of star and planet formation, perhaps throughout the Galaxy.
 
The standard model for the Earth's composition is that the planet formed  from material less volatile than sodium, but with an additional contribution of volatiles (Waenke \& Dreibus 1988).  
In this approach, carbon is considered a volatile element since, in thermodynamic equilibrium, in a mixture with a solar composition, it forms molecules such as CO and CH$_4$.  Therefore, the standard cosmochemical model presumes that the inner solar system was entirely vaporized at a temperature greater than 1,800 K (Grossman 1972).  However, the computed disk midplane temperature at 1 AU is less than 1,000 K in most models of accreting pre-main-sequence stars (D'Alessio et al. 2005).  Thus, primordial carbon grains should have been incorporated into the Earth.    Carbon must have been present in the nebular gas in a volatile form (CO, CH$_4$, or CO$_2$).  At 1 AU this gas would not be assimilated into silicate-rich planetesimals, but would be accreted by the Sun and gas giants or lost to space.   Beyond the nebular snow-line large bodies could incorporate the carbon grains and ices, which can be contributed to the young Earth by impacts.
Some previous solutions to this problem include ultraviolet heating and vaporization of infalling grains (Simonelli et al. 1997), reactions with OH (Gail 2002), and impacts (Bond et al. 2009).   However, the infall phase is short and models suggest that either none or all grains, including silicates, are vaporized (Simonelli et al. 1997) and in the warm inner disk OH may not be abundant enough (Glassgold et al. 2004; Glassgold et al. 2009; Woitke et al. 2009; Gorti \& Hollenbach 2004).

In this letter we meld measurements of relative amounts of carbon and silicon found within solar system bodies with our current understanding of the physics and chemistry in extra-solar protoplanetary systems to demonstrate that the innermost regions of the Solar Nebula was on ÒfireÓ and rocks forming within the terrestrial planet zone for the Sun and, likely other stars, are carbon-poor.

\section{Model Calculations and Results}

The paradigm of terrestrial planet formation involves the growth of solids in the solar nebula, which is a protoplanetary disk. Grains initially present in the disk upper layers settle to a dust-rich midplane where they coagulate and grow into planetesimals (Weidenschilling \& Cuzzi 1993). Depending upon their size, grains can be suspended in the disk atmosphere
via interactions with the turbulent gas (Weidenschilling \& Cuzzi 1993). In this theory sub-micron sized grains are held in the upper layers, by turbulent stirring, and coagulate to larger sizes where they eventually settle to the midplane. Grains settle to a vertical layer, $z_{sett}$, that depends on the disk physical structure, the strength of turbulence, parameterized by an $\alpha$ viscosity (Hartmann 2001), and the grain mass to surface area ratio. This height can be determined by setting the timescale of settling, $t_{sett}$, equal to the timescale of diffusive turbulent stirring of the disk, $t_{stir}$ (Dullemond \& Dominik 2004):

\begin{eqnarray}
\xi t_{sett} &=& t_{stir}\\
\xi\left[\frac{4}{3}\frac{\sigma}{m}\frac{c_s(z)}{\Omega_K^2}\rho(z)\right] &=&
\frac{{\rm Sc}(z)}{\alpha c_s(z)} \frac{z^2}{H_p}
\end{eqnarray}

\noindent $Sc$ is the Schmidt number which relates the decoupling of grains from turbulence.  $m/\sigma$ is the grain mass-to-cross section ratio, which is $m/\sigma = 4\rho_d a/3$, where $\rho_d$ is the grain density and $a$ is the grain radius. Here,  $\rho_d$ is calculated from graphite, whose density is 2.24 g cm$^{-3}$, with assumed porosities (Ormel et al. 2007).
$H_p$ is the pressure scale height which is given by $H_p = {c_{s,0}}/\Omega_k$.
$\Omega_K=\sqrt{G~M_\star/R^3}$ denotes the local Kepler
frequency and the sound speed at  the midplane, $c_{s,0} = \sqrt{\frac{kT_{gas}}{\mu_{gas}m_p}}$.
For perfectly coupled particles $Sc = 1$, which is the case for grains smaller than $a = 100$ $\mu$m (Dullemond \& Dominik 2004).
 When $\xi = 1$ the settling layer ($z_{sett}$), where grains are suspended by turbulence, is reached.
The depletion layer ($z_{depl}$), above which the disk is entirely depleted (by settling) of grains of that size, can be determined by setting $\xi = 100$ (Dullemond \& Dominik 2004).  
We adopt $\alpha=0.01$, which seems appropriate according to the Balbus-Hawley
instability mechanism to generate viscosity (Hartmann 2001).

The gas and dust temperature and density in the disk are calculated
by using the parameters listed in the caption of Figure 2 (see Nomura
et al. 2007 for details).\footnote{The gas heating via PAHs is included in the
model (e.g., Aikawa \& Nomura 2006), but we note that the temperature 
structure may have some dependency on models (e.g., Gorti \& Hollenbach
2004).}
For a dust model, we use a dust size distributions of $dn/da \propto
a^{-3.5}$ with grain radii ranging from 0.01 $\mu$m to 10 $\mu$m,
which are larger than those of the ISM grain model (Mathis et al.
1977), assuming the possibility of grain growth in dark clouds.
In the calculation of stirring depths, we adopt various porosities for
different grain sizes according to Ormel et al. (2007).

Figure 2 presents the estimates of $z_{sett}$ and $z_{depl}$ and the
gas and dust temperature profiles at 1 AU in our disk model. We also
show the chemical abundance profiles of key oxygen reservoirs\footnote{The 
chemical/thermal model did not include the chemistry of 
water and OH formation.  For these abundances we used the analytical 
approximations of Bethell \& Bergin (2009), which rely on knowledge of 
the local radiation field and gas temperature and density. The result is
consistent with other self-consistent model results (e.g. Gorti \& Hollenbach
2004).}.
Disk accretion onto the star and an active chromosphere
power X-ray and ultraviolet radiation fields that heat the
disk surface to high temperatures at the depth where grains are
suspended by turbulent stirring (Glassgold et al. 2004; Nomura et al.
2007); in contrast the disk midplane is cooler.
A key facet of disk thermal models is that gas and dust temperatures
are decoupled in the upper layers. 
Figure 2 shows that $T_{gas}$ between $z_{sett}$ and $z_{depl}$
can be quite high ($>500$K) for grains $\leq 10 \mu$m.
In addition, within this depth range oxygen-bearing
molecules are predominantly photodissociated with $n_O = 1.8 \times 10^{-4}n_H$ 
at $<$ 5 AU.
At 10 and 20 AU, however, the dust temperature is lower than the water 
desorption temperature (110 K) at all depths so that a significant amount 
of oxygen is locked on grain surfaces, as expected beyond the snowline.

Hot oxygen atoms erode carbon grains and release the carbon into the gas.
The physical and chemical conditions between $z_{sett}$ and $z_{depl}$ at the 
terrestrial planet forming zone in our model are such that the chemical
sputtering of carbon grains by hot oxygen atoms will be operative. Thus, for carbon, the need to
increase the dust temperature above the sublimation point, as in the
standard cosmochemical model, is evaded.    Silicates will not be subject to similar chemisputtering
(Barlow \& Silk 1977) and remain the seeds for terrestrial worlds.

The oxidation rate of carbon-based grains is: 
$k_{oxy}= n_O\sigma v_O Y$ (s$^{-1}$). 
$n_O$ and $v_O$ are the number density and the thermal velocity of oxygen 
atoms, respectively. $Y$ is the yield  (carbon atoms removed per incident 
oxygen atoms). We adopt laboratory studies for the yield as given in Table 2 
of Draine (1979), which covers the appropriate temperature range.
 For $T_{gas}=500$ K and 1000 K, $Y\sim 0.02$ and 0.2, respectively.
A summary of recent experiments at $T_{gas}>$ 1000 K can be found in Vierbaum and Roth (2002); these results
are consistent with our assumptions.

 The key question is the destruction timescale, $t_{dest}$, 
as grains cascade to smaller sizes when carbon atoms are removed.  
If $t_{dest}$ is shorter than $t_{stir}$, any carbon grain that is lifted 
above the settling layer will be destroyed.  
To determine $t_{dest}$ we have constructed a kinetic chemical network, 
such as $g_{n+1} + O  \rightarrow g_n + C + O$ (with n = \# of carbon atoms) 
with the rate specified above.
This network is then extended down to carbon atoms and we solve the system 
of ordinary differential equations to calculate $t_{dest}$ of the biggest 
grain (0.002 $\mu$m) feasible to the computational memory and time. 
We then extend these results to larger grains by extrapolating this timescale
(see Eq. (4) below). 

As shown above, the reaction rate ($k_{oxy}$) is proportional to the grain 
surface area ($\sigma=\pi a^2$) in a given temperature and abundance of atomic 
oxygen.
For $\sigma$, we have related the grain volume to its mass:

\begin{equation}
\frac{4}{3}\pi a^3 = \frac{m_C}{\rho_d},
\end{equation}

\noindent Here $m_C$ is the total mass of carbon atoms contained in the given 
size grain. Based on this model with no porosity, a 0.002 $\mu$m grain 
($g_{3000}$) contains 3000 carbon atoms.
For the initial abundance of each grain ($g_\#$) in this kinetic calculation, 
we adopted the total number of carbon atoms (per H atom) derived from the MRN 
interstellar grain size distribution for graphite (see Figure 6.19 of Tielens 
2005) and extrapolated to very small grains.  

Figure 3 shows the abundance variation of atoms in carbon grains as a function 
of time at 1 AU. The physical conditions for the kinetic calculation are 
obtained from our disk model at the depth (0.122 AU) where $T_{gas}=500$ K. 
These results demonstrate that, assuming no source term, there is a rapid decay of carbon grains
to reduced carbon content and size.
We define $t_{dest}$ of a certain size of grain when $X(g_1)$ decreases 
by 4 orders of magnitude in the reservoir. 
Therefore, 0.002 $\mu$m grain ($g_{3000}$) is 
destroyed within 2 years at the depth of $T_{gas}=500$ K at 1 AU. 
$t_{dest}$ of 0.002 $\mu$m grain at 3, 10, and 20 AU are 14, 100, and 400 
years, respectively (see Table 1). 
In the calculation of $t_{dest}$ at 10 and 20 AU, we assume all oxygen is 
atomic although a significant amount is likely locked on grain 
surfaces as water ice. 
Therefore, the calculated $t_{dest}$ at 10 and 20 AU is
the lower limit.
In this disk model, the same size grains 
are destroyed at the depth where $T_{gas}=1000$ K about 3 times faster 
compared to the depth where $T_{gas}=500$ K.   Our estimates of destruction timescales
are very conservative as we have examined the timescale at 500 K and defined the timescale
after a several orders of magnitude decrease in abundance.

As grains grow in size by colliding aggregation, the porosity, the measurement 
of volume that is not occupied by carbon atoms, increases (Ormel et al. 2007).  
The enlargement parameter $\psi$ is defined by $\psi=V/V^*$, where $V^*$ is the 
volume that carbon grains occupy in its compacted state while $V$ is the total 
spatial extent of the given size grain.    Thus, the porosity is defined as 
$[(V-V^*)/V]\times100=[1-1/\psi]\times100$. Therefore, the same size of grain 
of a higher porosity contains a lower number of carbon atoms to reduce 
$t_{dest}$. $\psi=$1 does not necessarily mean no porosity, but we assume no 
porosity for $\psi=$1 for the conservative calculation of $t_{dest}$ (i.e., 
the longest destruction timescale) of a given size grain. According to the 
results of Ormel et al. (2007), we adopt $\psi=$1, 8, and 27 for 0.1, 1, and 
10 $\mu$m grains, respectively. Therefore, the porosity of 0.1, 1, and 10 
$\mu$m grains are 0, 88, and 96\%, respectively. 

$t_{dest}$ at a given condition is proportional to 
$1/k_{oxy} \times \#(C)_{grain}$, where $\#(C)_{grain}$ is the total number of 
carbon atoms contained in the grain, and thus, proportional to the size of 
grain ($1/\sigma \times a^3 \propto a$). 
However, the porosity reduces $t_{dest}$ since $\#(C)_{grain}$ of a given 
size decreases. 
Therefore, $t_{dest}$ of a given size grain can be calculated by:

\begin{equation}
 t_{dest, a}=(\frac{a}{0.002~{\rm \mu m}})\psi_{a}^{-1}~t_{dest, 0.002}, 
\end{equation}

\noindent where $t_{dest, 0.002}$ and $t_{dest, a}$ are $t_{dest}$
of 0.002 $\mu$m grain and a grain with size $a$.    $\psi_{a}$ is the enlargement
factor for grains of a certain size.
The results of the combination of grain size and porosity are presented in 
Table 2. At the depth of $T_{gas}=500$ K, $t_{dest} < t_{stir}$ for grains 
smaller than 10 $\mu$m at 1 and 3 AU while at 10 and 20 AU, only very small 
grains (0.01 $\mu$m) satisfy the condition. Therefore, all primordial carbon 
grains are destroyed within the terrestrial planet forming zone by turbulent 
stirring above $z_{sett}$ within $t_{stir}$.   
This effectively stops incorporating carbon grains in the initial stages of 
planetesimal formation -- the coagulation of small grains in the upper layers 
and the subsequent settling to lower heights. 
Moreover, this mechanism will decrease in efficiency within the asteroid belt, 
which could produce the observed asteroid compositional gradients.

\section{Discussion}

Carbon grain destruction in the dense UV free midplane has been explored by  Finocchi et al. 1997, Bauer et al. 1997, and Gail (2001, 2002).    In this work OH is the primary agent for carbon grain destruction.   However, the amount of OH in the disk midplane has some uncertainty as these authors determine an OH abundance of $\sim 10^{-10} - 10^{-7}$ relative to H$_2$ (Finocchi et al. 1997, Bauer et al. 1997), while Glassgold et al. (2009) find an abundance well below $10^{-12}$.    Our model relies on an abundance of oxygen atoms on the disk surface (e.g. Fig.~2) in a region that is clearly exposed to UV radiation, ensuring a ready supply of atoms
with O/OH $\gg$ 1.   However, our model does not preclude destruction of carbon grains in the midplane by OH, if the abundance is high enough.

As a result of chemical destruction of carbon grains all carbon in the inner solar system will reside in the gas, which will accrete into the star or dissipate.  Carbon
grains that reach sizes of $a > 50-100$ $\mu$m in the outer disk will settle below the layer where chemisputtering dominates.   This brings up an important question: if the Earth did not incorporate carbon grains or accrete sufficient amounts from the gas, how did the Earth assimilate any carbon?  We suggest that maybe the problem of the Earth's carbon  
is related to the origin of the Earth's oceans.   At 1 AU it is likely that grains were too hot to be sufficiently coated with ices and instead the Earth potentially received its water from impacts of bodies that originate beyond the nebular snow-line near 2.5 AU (e.g. Morbidelli et al. 2000, but see also Drake 2004).   These bodies are likely also carbon rich (e.g. Gradie \& Tedesco 1982) and could have brought carbon with water ice.     The mass of  carbon in the Earth's mantle is $\sim 5.5 \times 10^{23}$ g with perhaps up to an order of magnitude additional carbon sequestered in the core (All\'egre et al. 2001).   In the study of terrestrial planet formation performed by  O'Brien et al. (2006) they estimate that $\sim 15$\% (median value of simulations) of the Earth was delivered by bodies that originated beyond 2.5 AU.  If we assume that these bodies contained the same amount of carbon as in carbonaceous chondrites of $\sim 10$ mg/g (Waenke \& Dreibus 1988) then this corresponds to a delivery of roughly 10$^{25}$ g of carbon, which is in the exact range of that estimated for the Earth (see Bond et al. 2009 for an additional look at this issue).

\acknowledgments
We are grateful to M. Jura for motivating this work and J. Cuzzi for helpful discussions. 
This work was supported by the National Research Foundation of Korea(NRF) grant 
funded by the Korea government(MEST) (No.2009-0062865).

\clearpage

\begin{figure}
\figurenum{1}
\epsscale{1.0}
\plotone{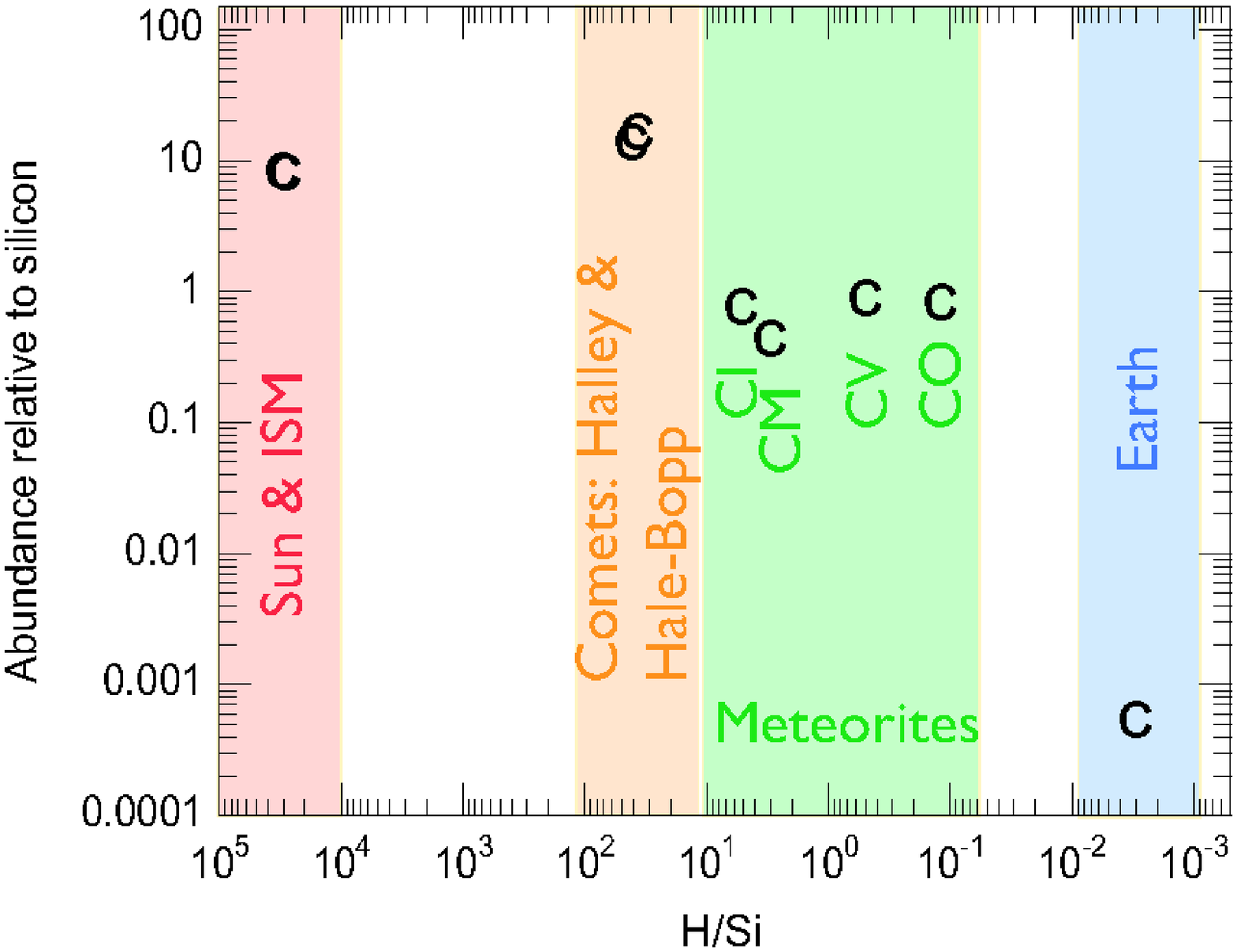}
\caption{
Carbon abundances relative to Silicon in the Solar System. Abundance ratios of carbon (listed as C) relative to silicon for representative bodies in the solar system as a function of their H/Si content. Shaded areas denote specific types of bodies (2 Oort cloud comets, 4 classes of chondritic meteorites, Earth) or instances with strong similarities (Sun and ISM). Abundance estimates are taken from the literature (Geiss 1987, Wasson \& Kallemeyn 1988, Sofia et al. 1994, All\'egre et al. 2001, Min et al. 2005, Grevesse et al. 2007). The amount of hydrogen on the Earth is estimated assuming a hydrosphere mass of $1.4\times10^{24}$ g.   Figure adapted and expanded from Geiss 1987.
}
\end{figure}

\clearpage

\begin{figure}
\figurenum{2}
\epsscale{1.0}
\plotone{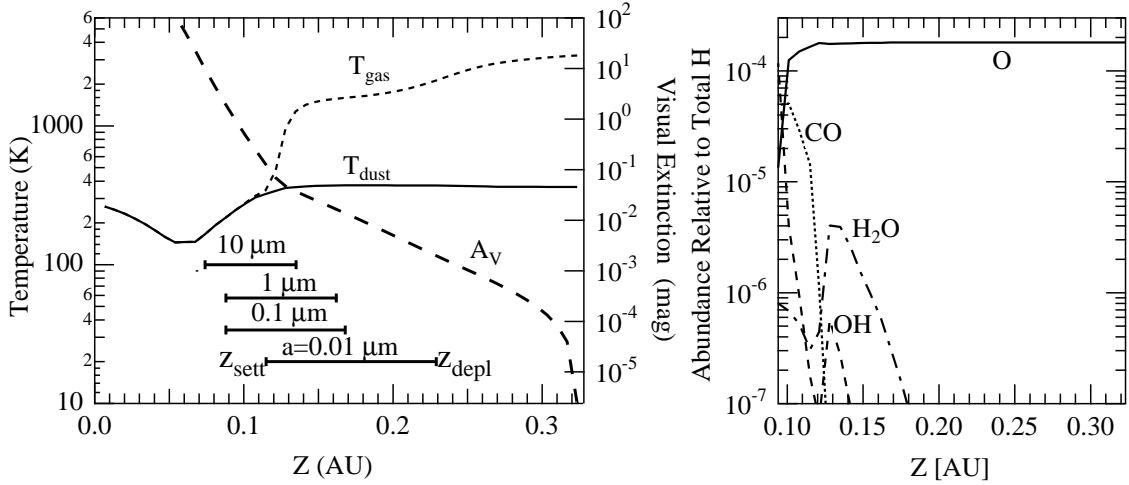}
\caption{ {\em Left:}   Temperatures and $z_{sett}$ and $z_{depl}$ in the protoplanetary disk atmosphere.  
Gas and dust temperatures as a function of vertical height above the midplane at 1 AU. Also shown are the estimated depletion and settling depths for various grain sizes. {\em Right:} Chemical abundances of key oxygen carriers as a function of vertical depth.   The chemical calculation is stopped at $z \sim 0.1$ AU when the gas/dust temperatures reach equilibrium.
The model assumes an axisymmetric $\alpha$-viscous disk (e.g.,
Hartmann 2001) with $\alpha=0.01$ and $\dot{M}=10^{-8}M_{\odot}$ 
yr$^{-1}$, which is surrounding a central star with parameters of
$M_*=1.0M_{\odot}$, $R_*=2R_{\odot}$, and $T_*=4000$K.  Both UV and X-ray spectral profiles from the central star are assumed to be the same as those of TW Hydra, and they are normalized so that the total FUV luminosity is $L_{FUV}=1\times10^{31}$ ergs s$^{-1}$ (which corresponds to $G_{FUV} =300~ G_0$ at the disk radius of 100 AU, where $G_0=1.6\times10^{-3}$ ergs s$^{-1}$ cm$^{-2}$) and the X-ray luminosity is $L_X=1\times10^{29}$ ergs s$^{-1}$.}
\epsscale{1.0}
\end{figure}

\clearpage

\begin{figure}
\figurenum{3}
\epsscale{1.0}
\plotone{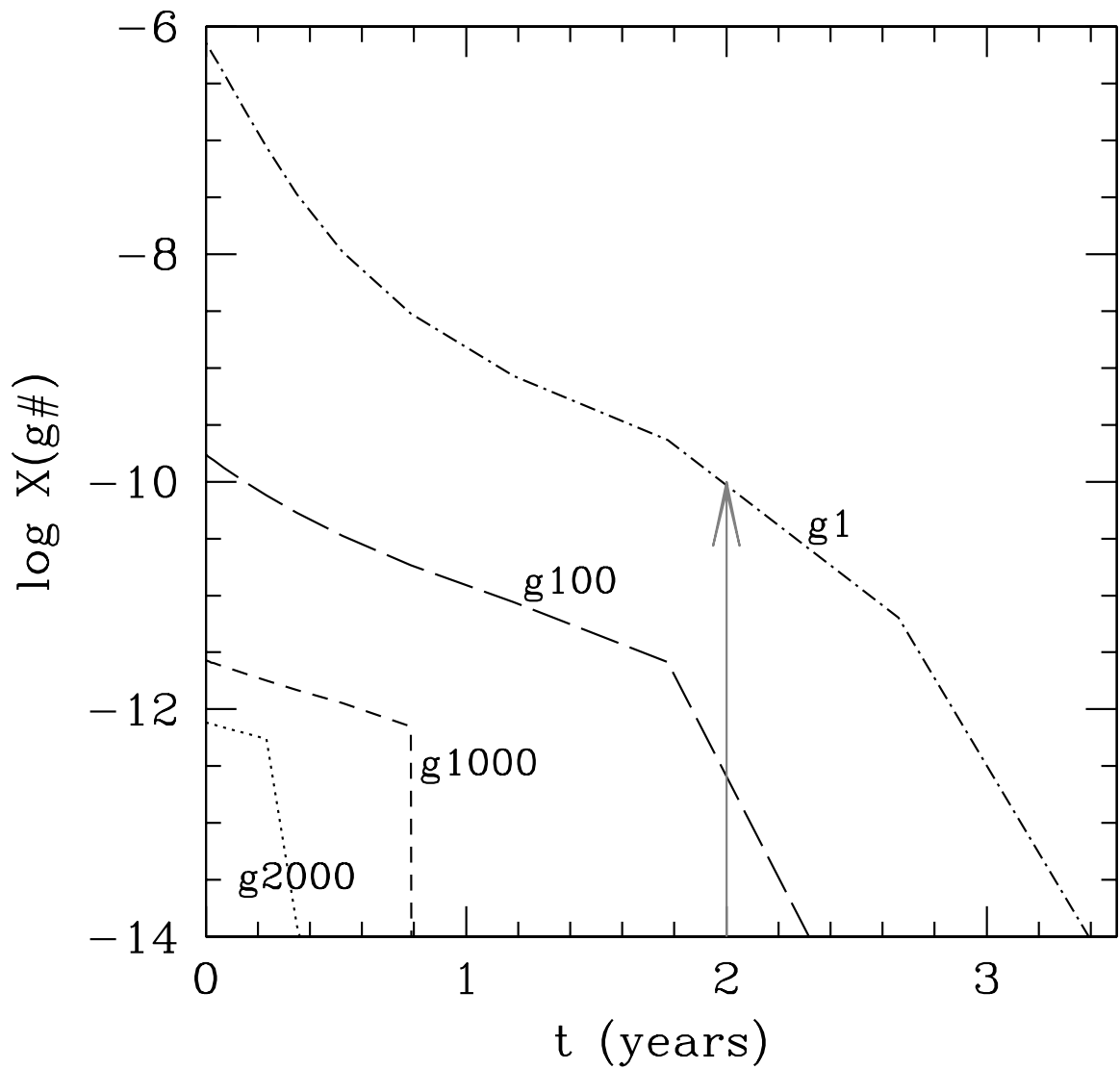}
\caption{
Kinetics of carbon grain destruction at 1 AU. Plot of grain abundances -- maximum grain size ($a(g_{3000})$) is $\sim 0.002$ $\mu$m -- with variable number of carbon atoms as a function of time using conditions 
at the depth where $T_{gas}=500$ K (see Table 1). Various line types represent grains containing different carbon atoms. The gray arrow indicates the timescale ($t_{dest}$ of 0.002 $\mu$m grain) in which the abundance of $g_1$ decreases by 4 orders of magnitude.}
\end{figure}

\clearpage

\begin{table}
\centering
\caption{Destruction timescale of 20~\AA~ grains at the depth of $T_{gas}=500$ K}
\begin{tabular}{lcccc}
\hline\hline
\multicolumn{1}{c}{} &
\multicolumn{1}{c}{1 AU} &
\multicolumn{1}{c}{3 AU} &
\multicolumn{1}{c}{10 AU} &
\multicolumn{1}{c}{20 AU} \\\hline
Depth (AU) & 0.122 & 0.494 & 2.047 & 5.180 \\
H$_p$ (AU) & 0.033 & 0.101 & 0.460 & 1.164 \\
n(H$_2$) (cm$^{-3}$) & $1.42\times 10^9$ & $2.03\times 10^8$  & $2.50\times 10^7$ &$7.50\times 10^6$   \\
$t_{stir}$ (yr) & 160 & 840 & 3200 & 8400 \\
$t_{dest}$ (yr) & 2 & 14 & 100 & 400 \\\hline
\end{tabular}
\label{tab:timesc1}
\end{table}

\begin{table}
\centering
\caption{Destruction timescale of various size of grains at the depth of $T_{gas}=500$ K}
\begin{tabular}{cccccc}
\hline\hline
\multicolumn{1}{c}{Grain Size ($\mu$m)} &
\multicolumn{1}{c}{Porosity (\%)} &
\multicolumn{1}{c}{1 AU} &
\multicolumn{1}{c}{3 AU} &
\multicolumn{1}{c}{10 AU} &
\multicolumn{1}{c}{20 AU} \\
\multicolumn{1}{c}{} &
\multicolumn{1}{c}{} &
\multicolumn{1}{c}{(160 yr)\tablenotemark{a}} &
\multicolumn{1}{c}{(840 yr)\tablenotemark{a}}&
\multicolumn{1}{c}{(3200 yr)\tablenotemark{a}} &
\multicolumn{1}{c}{(8400 yr)\tablenotemark{a}} \\\hline
0.01 & 0 & 10 yr & 70 yr & 500 yr & 2000 yr   \\
0.1 & 0 & 100 yr & 700 yr & 5000 yr & 20000 yr \\
1 & 88 & 125 yr & 875 yr & 6250 yr & 25000 yr \\
10 & 96 & 370 yr & 2593 yr & 18520 yr & 74000 yr \\\hline
\end{tabular}
\label{tab:timesc2}
\tablenotetext{a}{$t_{stir}$ at each radius as given in Table 1.}
\end{table}


\begin{references}
\noindent
\reference{} Aikawa, Y. \& Nomura, H. 2006, ApJ, 642, 1152
\reference{} All\'egre, C.,  Manh\'es,  \& G., Lewin, E. 2001, Earth and Planetary Science Letters, 185, 49
\reference{} Barlow, M.~J., \& Silk, J.\ 1977, \apj, 215, 800 
\reference{} Bauer, I., F.~Finocchi, W.~J.~Duschl, H.-P.~Gail, and J.~P.~Schloeder 1997, A\&A, 317, 273
\reference{} Bethell, T. J. \& Bergin, E.A., Science, in press
\reference{} Bond, J.~C., Lauretta, D.~S., \& O'Brien, D.~P.\ 2009, Icarus, in press
\reference{} D'Alessio, P., Calvet, N., \& Woolum, D. S. 2005, ASPCS, 341, 353
\reference{} Draine, B. T.  Astrophys. J., 1979, 230, 106-115
\reference{} Drake, M.~J.\ 2004, M\&PS Supp., 39, 5031 
\reference{} Dullemond, C. P. \& Dominik, C. 2004 A\&A, 421, 1075 
\reference{} Finocchi, F., H.-P.~Gail, and W.~J.~Duschl 1997, A\&A, 325, 1264 
\reference{} Gail,  H.-P. 2002, A\&A, 390, 253
\reference{} Gail, H.-P.\ 2001, A\&A, 378, 192 
\reference{} Geiss, J. 1987, A\&A., 187, 859
\reference{} Glassgold, A.~E., R.~Meijerink, and J.~R.~Najita 2009, ApJ, 701, 142 
\reference{} Glassgold, A. E., Najita, J., \& Igea, J. 2004,  ApJ., 615, 972
\reference{} Goto, M., and 10 
colleagues 2009, \apj 693, 610 
\reference{} Gorti, U. \& Hollenbach, D. 2004, 613, 424
\reference{} Gradie, J. \& Tedesco, E. 1982, Science, 216, 1405
\reference{} Grevesse, N., Asplund, M., \& Sauval, A. J. 2007, Space Science Reviews, 130, 105
\reference{} Grossman, L. 1972, Geochim. Cosmochem. Acta, 36, 597
\reference{} Hartmann, L. 2001, in Accretion Processes in Star Formation (Cambridge Univ. Press), pp. 95-96
\reference{} Jura, M. 2006, ApJ., 653, 613
\reference{} Jura, M. 2008, AJ., 2008, 135, 1785
\reference{} Mathis, J. S., Rumpl, W., \& Nordsieck, K. H. 1977, ApJ., 217, 425 (MRN)
\reference{} Min, M., Hovenier, J. W., \& de Koter, A., Waters, L. B. F. M., \& Dominik, C. 2005, Icarus, 179, 158
\reference{} Morbidelli, A., J.~Chambers, J.~I.~Lunine, J.~M.~Petit, F.~Robert, G.~B.~Valsecchi, and K.~E.~Cyr 2000,  M\&PS, 35, 1309
\reference{} Nomura, H., Aikawa, Y., Tsujimoto, M., Nakagawa, Y., \& Millar, T. J., 2007, ApJ., 661, 334
\reference{}  O'Brien, D.~P., A.~Morbidelli, and H.~F.~Levison 2006, Icarus 184, 39 
\reference{} Ormel, C. W., Spaans, M., \& Tielens, A. G. G. M. 2007, A\&A., 461, 215
\reference{} Paquette, C., Pelletier, C., Fontaine, G., \& Michaud, G. 1986, ApJ., 61, 197
\reference{} Savage, B. D. \& Sembach, K. R. 1996, Ann. Rev. Astron. Astrophys., 34, 279
\reference{} Simonelli, D. P., Pollack, J. B., \& McKay, C. P. 1997, Icarus, 125, 261
\reference{} Sofia, U. J., Cardelli, J. A., \& Savage, B. D. 1994, ApJ., 430, 650
\reference{} Tielens, A. G. G. M. 2005, in The Physics and Chemistry of the Interstellar Medium (Cambridge Univ. Press), pp. 219
\reference{} Vierbaum, R. \& Roth, P. 2002, Proceedings of the Combustion
Institue, 29, 2423
\reference{} Wasson, J. T., Kallemeyn, G. W. 1988, Royal Society of London Philosophical Transactions Series A, 325, 535
\reference{} Waenke, H. \& Dreibus, G. 1988, Royal Society of London Philosophical Transactions Series A 325, 545
\reference{} Weidenschilling, S. J. \& Cuzzi, J. N. 1993, in Protostars and Planets III, 1031
\reference{} Woitke, P., Kamp, I. \& Thi, W.-F. 2009, A\&A, 501, 383
\end{references}
\end{document}